\documentclass[twoside,twocolumn]{article}

\usepackage{blindtext} 

\usepackage[sc]{mathpazo} 
\usepackage[T1]{fontenc} 
\usepackage{microtype} 

\usepackage[english]{babel} 

\usepackage[hmarginratio=1:1,top=32mm,columnsep=20pt]{geometry} 
\usepackage[hang, small,labelfont=bf,up,textfont=it,up]{caption} 
\usepackage{booktabs} 

\usepackage{amsmath}

\usepackage{enumitem} 
\setlist[itemize]{noitemsep} 

\usepackage{abstract} 

\usepackage{titlesec} 
\renewcommand\thesection{\Roman{section}} 
\renewcommand\thesubsection{\roman{subsection}} 
\titleformat{\section}[block]{\large\scshape\centering}{\thesection.}{1em}{} 
\titleformat{\subsection}[block]{\large}{\thesubsection.}{1em}{} 

\usepackage{graphicx}
\graphicspath{ {/Users/johnnyseales/Desktop/Active_Projects/NASA_search_strategy/Drafts/Conversion/Revision} }

\usepackage{fancyhdr} 
\usepackage{titling} 

\usepackage{hyperref} 

\pretitle{\begin{center}\Huge\bfseries} 
\posttitle{\end{center}} 
\title{Different is More: The Value of Finding an Inhabited Planet that is Far From Earth2.0} 
\author{%
\textsc{Adrian Lenardic}\thanks{ajns@rice.edu} \\[1ex] 
\textsc{Johnny Seales} \\[1ex] 
\normalsize Rice University \\ 
}
\date{\today} 
\renewcommand{\maketitlehookd}{%
\begin{abstract}
\noindent The search for an inhabited planet, other than our own, is a driver of planetary exploration in our solar system and beyond. Using information from our own planet to inform search strategies allows for a targeted search. It is, however, worth considering some span in the strategy and in \emph{\emph{a priori}} expectation. An inhabited, Earth-like planet is one that would be similar to Earth in ways that extend beyond having biota. To facilitate analysis, we employ a metric akin to the Earth-similarity index of Schulze-Makuch et al. [2011]. The metric extends from zero, for an inhabited planet that is like Earth in all other regards (i.e., zero differences), toward end-member values for planets that differ from Earth but maintain life potential. The analysis shows how finding inhabited planets that do not share all other Earth characteristics could improve our ability to assess galactic life potential without a large increase in time-commitment costs. Search strategies that acknowledge the possibility of such planets can also minimize the potential of exploration losses (e.g., searching for long durations to reach conclusions that are search strategy biased). Discovering such planets could additionally provide a test of the Gaia hypothesis - a test that has proved difficult using only the Earth as a laboratory. Lastly, we discuss how an Earth2.0 narrative that has been presented to the public as a search strategy comes with nostalgia-laden philosophical baggage that does not best serve exploration.
\end{abstract}
}

\begin{document}

\maketitle



\section*{Introduction}

The idea of a second Earth has a long history [Couprie, 2011]. Recently, NASA has entered into second Earth thinking: 1) ``This discovery gives us a hint that finding a Second Earth is not a matter of if but when.'' - Thomas Zurbuchen, Assoc. Admin., Science Mission Directorate at Nasa, 2017; 2) ``This exciting result brings us one step closer to finding an Earth 2.0.'' - John Grunsfeld, Assoc. Admin., Science Mission Directorate at Nasa, 2015. The quotes come from press conferences that announced discoveries regarding planets orbiting stars other than our own (exoplanets). A motivator behind these statements is the search for life beyond Earth and placing observational constraints on the probability of life in our galaxy.  

Debates about life beyond Earth have a long history. An even-handed referencing would take pages. Interested readers can easily track down a multitude of books, articles, blogs, and the like. As a starting point, two books that encapsulate competing views are Ward and Brownlee [2000] and Kasting [2010]. End-member views regarding galactic life potential are ``rare-Earth'' and ``plenitude'' (i.e. life requires specific environmental conditions, of the kind that exist on Earth, versus the idea that life can thrive in a range of conditions that might exist across planets and/or moons within our galaxy). To be clear (and avoid strawmen): 1) Rare does not mean singular (a rare-Earth view does not exclude life beyond Earth - it posits that in order for a planet to have life, particularly complex life, it must share certain essential features with the Earth, beyond having life, that make the Earth the planet it is and, it argues, that the combination of such features is rare for planets in our galaxy); 2) Plentitude does not mean all things are possible (a plentitude view acknowledges that there will be galactic bodies with conditions that do not allow for life); 3) Nature need not care about end-members (as an example: conditions that make the Earth the Earth, beyond having life, may not be rare in the galaxy such that there are many planets that are essentially analogous to Earth; life may only exist on such planets which would not be in line with a plentitude end-member view and, although it would not support the idea that Earth conditions are rare, it would confirm one aspect of the rare-Earth argument, i.e., Earth conditions are needed for life). 

We are now at the stage of exoplanet research where missions are being designed with the goal of detecting remote signatures of planetary life [e.g., Schwieterman et al., 2018]. This is a new phase of exploration into galactic life potential. It generates a level of excitement that motivated the press release statements quoted in the first paragraph. It brings with it the potential that observational data can be brought to bear on long-standing debates about galactic life. It also brings with it an equally old question related to any exploration: how will we choose to look? This is a decision problem. It is a decision problem under uncertainty. Any exploration comes with risk. Uncertainty adds to that risk. Acknowledging uncertainty and risk does not damp excitement. However, the fact that a particular search strategy is gaining traction, one that is centered around Earth analogs, without being weighed against alternates in light of uncertainty and risk, does motivate the step back and rethink that is the core of this paper (note added in revision: over the time that this paper was in review we became aware that we are not the first to suggest a rethink regarding search strategy [Bean et al., 2017; Kite et al., 2018; Lingam and Loeb, 2018a]). 

A search strategy can be centered on planets that share a high number of Earth attributes, with the thought that this maximizes our chances of finding signs of life. Our central thesis is that a search strategy that allows for deviations from Earth analogs can bring more information and less risk at moderately higher cost. Fleshing this statement out forms the bulk of this paper. 

There is an added reason why we argue for a break from an Earth centered view. Wanting to know if life exists beyond Earth, and what that implies for life potential across our galaxy, is not driven by scientific motivations alone. The humanistic implications of the search, and the humanistic/cultural factors that feed into it, should be acknowledged. More specifically, the desire to find a planet like our own is driven by factors that extend well beyond pure scientific curiosity [Messeri, 2016]. Our intent is not to argue that humanistic/cultural aspects be removed from the discussion. Rather, we want to layout how finding inhabited planets that do not look like home to us could carry broader scientific and humanistic implications than finding an Earth2.0. The humanistic/philosophical implications are discussed in the final discussion section.


\section*{Non-Earthness, Planetary Life Potential, and Search Strategies}

Exploration goal(s) determine search strategies. The goal(s) of exoplanet exploration, as related to galactic life, can be expressed as a single question or intertwined questions. There is an existence question: `does life exist beyond Earth?' If this is the only motivating question, then the search endeavor could be of the `find one and done' type. We suspect that most people who have dedicated time and thought to exoplanet explorations do not subscribe to this. Thus, there must be a broader goal. We would argue that addressing the issue of galactic life potential (rare versus plentiful) is that broader goal. That introduces the intertwined questions of `what conditions allow for and/or maximize life potential (i.e., the probability that life exists on a planet)?' and `what percentage of planets in our galaxy are inhabited (i.e., what percentage have a biosphere)?' For what follows we will, unless otherwise noted, assume all of these questions are motivators in the collective search for life beyond Earth and, as such, all should be considered when determining search strategies. 

We now pose an added question: `would finding inhabited planets that share many other Earth attributes or finding inhabited planets that differ significantly from Earth better help us to address the question of galactic life potential and should this consideration feed into search strategies?' We will argue that it should be considered. This leads to a practical issue that needs to be addressed from the start: an overly broad search is not practical. Given that, we would like some constraints on how different we think a planet could be from Earth and still maintain life potential. This motivates the first sub-section below.

\subsection*{How Different can Different Be and Maintain Life Potential}

The term ``Earth-like'' is often used in a qualitative way. A quantitative metric can be defined as per the Earth-similarity index of Schulze-Makuch et al. [2011]. Our interest is more specific than asking how close a planet is to Earth in terms of the full range of variables that come into play. We are interested in the question of how different a planet can be from Earth and still allow for life. We can, for example, use the Earth-similarity index and define a sub-range over which a planet can maintain life potential (this can also be applied to moons). Since our interest is on how different a galactic body can be from the Earth and allow for life, it will be useful to center our metric on zero  (i.e., if a planet with life potential is like Earth in all other ways then the value of the metric is zero). However one defines such a metric a useful next step will be to ask what are the extreme values it can take while maintaining a non-negligible probability for life. 

We start with an agreed upon criteria: life requires energy. The energy sources for a biosphere are energy from the star a planet or moon orbits and internal energy from the decay of radioactive isotopes within its interior, heat retained from formation, and/or tidal heating.
 
Figure 1a is a schematic of how energy sources affect life on Earth. Solar and internal energy can be used as direct energy sources to power photosynthesis or chemosynthesis. The energy sources also drive cycles that influence environmental conditions. If there is a limited range of environmental conditions under which a biosphere can exist, this implies that energy sources can affect planetary life potential by maintaining livable conditions. For life as we know it, the existence of liquid water is crucial. The classic idea of a "Habitable Zone" ties into delineating conditions required for a planet to maintain liquid water over time scales that allow for life development and evolution [Kasting et al., 1993]. 

\begin{figure*}[!htb]
\includegraphics[width=\textwidth]{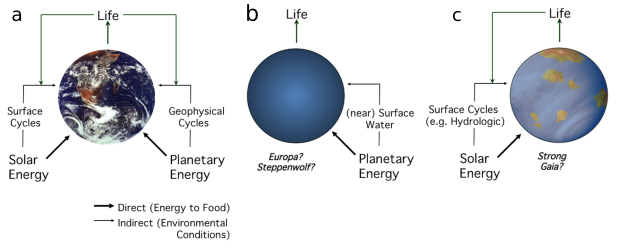}
\caption{Figure 1: a) Schematic of how energy sources affect life, directly and indirectly, on Earth. b) Schematic of potentially livable planets/moons on which solar energy is not critical to life. c) Schematic of potentially livable planets/moons on which internal energy is not critical to life.}
\end{figure*}

For the Earth, the buffering of environmental conditions is generally considered to rely on hydrological and geophysical/geochemical cycles. Internal energy drives volcanic and tectonic activity that transfers volatiles (CO$_2$, H$_2$O) between a planet's surface envelopes (atmosphere, hydrosphere, biosphere) and its rocky interior (crust, lithosphere, mantle). Volcanism cycles greenhouse gases into the atmosphere. Tectonics creates weatherable topography and weathering reactions draw greenhouse gases out of the atmosphere. Weathering depends on hydrology. This means that surface and deep planet cycles are linked in so far as discussions of buffering Earth's climate are concerned [Kump et al, 2000]. Life also links in as it has the ability to effect the cycles that maintain environmental conditions suitable for its own existence [Lovelock and Margulis, 1974]. Over geologic timescales, atmospheric CO$_2$ content and associated greenhouse climate forcing is influenced by the balance between volcanic degassing and weathering [Berner et al., 1983]. Weathering depends on processes governed partly by surface temperature, which allows for the potential that a planet can buffer/stabilize climate and surface conditions in a manner that allows liquid water to exist over extended time frames [Walker et al., 1981].   

The silicate-weathering negative feedback, outlined above, is the currently preferred hypothesis for how the Earth's climate has been regulated so as not to enter a prolonged hard snowball state or a runaway greenhouse state. This has cast a significant influence on ideas about how to search for inhabited planets beyond our solar system [e.g., Kasting, 2010]. It is worth stressing that, as formulated for Earth climate stabilization, the feedback relies on both solar and internal planetary energy. Removing one of the two energy sources thus affects direct energy sources for life and also a mechanism for maintaining environmental conditions conducive to life. This stresses how a planet or moon that lacks one of the two energy sources would be distinctly non Earth-like.  

The idea that a planetary body can have life even if solar energy is negligible (Figure 1b) is driving exploration of icy moons within our own solar system [Schultze-Makuch and Irwin, 2001; Wenz, 2017] and has been suggested for planets that do not orbit stars [Abbot and Switzer, 2011]. For planets that do not orbit stars the difficulty of remote life detection is extreme. For planets/moons that do orbit a star there is the potential that life on such bodies would not interact with an atmosphere so as to create biosignatures that can be observed remotely [Schwieterman et al., 2018]. However, it is not clear that life would not leave detectable signatures that are not connected to atmospheric chemistry [Lingam and Loeb, 2018b]. Detectability cannot be ignored but at this stage we are concerned about limits. What is key to our arguments is that an inhabited planet that lacks solar energy affecting life directly or indirectly remains a viable possibility. 

An inhabited planet that lacks internal energy sources (Figure 1c) would imply that geophysical/geochemical cycles are not required to maintain conditions conducive to life. Habitable conditions can be maintained on waterworlds (planets with water masses 10-1000 times that of Earth) as a result of stochastic variations in formation conditions with no need for geochemical cycling  [Kite and Ford, 2018]. The potential that habitable conditions can be maintained without geo-cycling can extend beyond waterworlds if life itself maintains conditions that allow for its continued existence. This is the core of Gaia theory [Lovelock and Margulis, 1974; Lovelock, 1979; 1995; Watson and Lovelock, 1983]. Over the course of debating the theory, different levels of Gaia have been suggested [Kirchner, 1989; 2003]. Soft Gaia considers life on Earth to have influence on the geophysical/geochemical cycles that modulate Earth's surface environment while the strong form of Gaia considers life to be critical to modulating surface conditions at livable levels [Barlow, 1991; Schneider et al., 2008].  Under a strong Gaia view, life could exist on a planet or moon that has tapped all of its internal energy. Akin to the discussion of the previous paragraph, the key for what follows is that an inhabited planet that lacks internal planetary energy affecting life directly or indirectly remains a viable possibility.

The scenarios of Figures 1b and 1c are energetic extremes. Between them sits the potential of livable planets that differ from the Earth in other ways. Some examples: Planets without oceans could allow for habitable conditions [Abe et al., 2011]; Planets that are ocean worlds could allow for life [Kaltenegger and Sasselov, 2011]; Planets with internal energy principally driven by tidal heating (a minor factor for Earth) could allow for life [Barnes et al., 2009]; Planets without plate tectonics (the geologic mode of Earth) could allow for conditions conducive to life [Lenardic et al., 2016b; Foley and Smye, 2018]. 

\begin{figure}[!htb]
\includegraphics[width=0.5\textwidth]{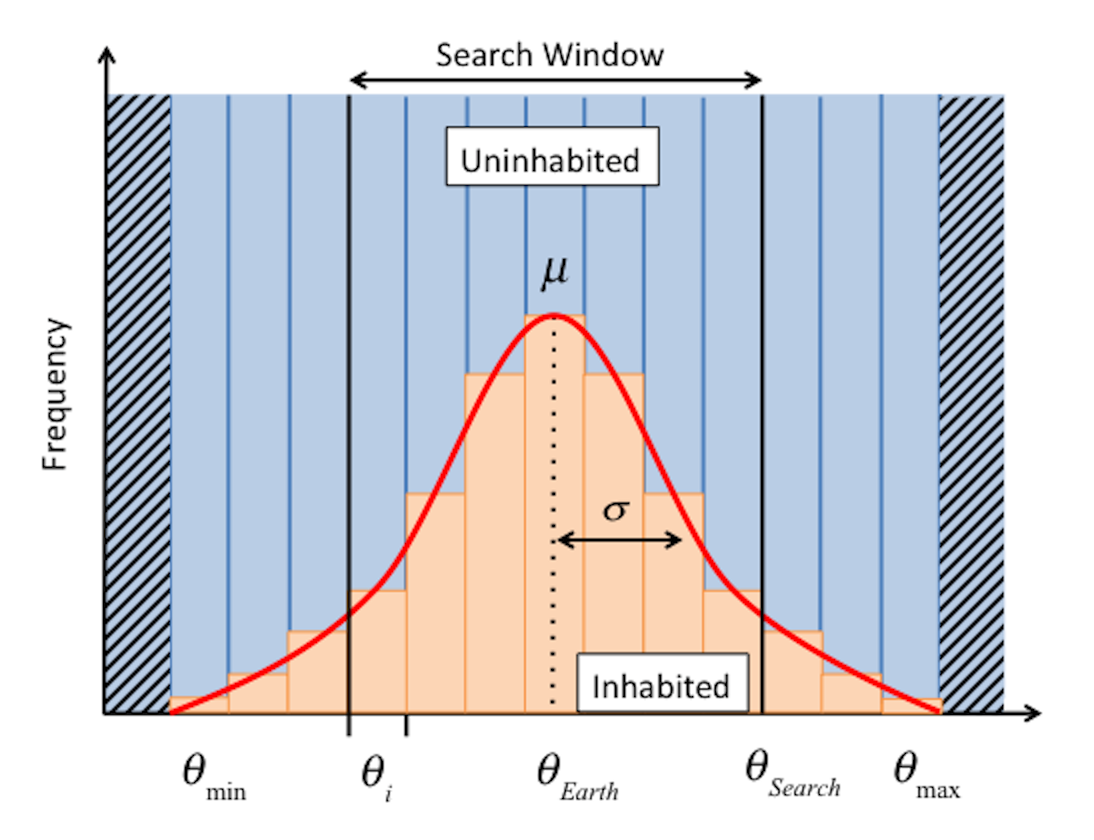}
\caption{Future observations, limited by a window of potential habitability ($\theta$) and by the search strategy employed, are grouped ($\theta_i$)  relative to Earth ($\theta_{Earth}$). Each group has a different habitability potential and there is a set of conditions that maximize habitability potential ($\mu$). The width of the search window ($\theta_{Search}$)  affects how accurately we can predict these conditions.  }
\end{figure}

\subsection*{Galactic Potentialities, Cost, Gain, and Risk: Statistical Thought Experiments}
In this sub-section, we explore the degree to which different search strategies can alter our ability to address questions of galactic life. The analysis will be in the form of double `what if' thought experiments. The distribution of life in the galaxy is unknown but different potentialities can be considered (e.g., `what if life potential is not peaked around Earth') to evaluate how different search strategies perform under variable possibility space (e.g., `what if a targeted search strategy is used under this potential life probability distribution'). The mathematical details of the experiments can be found in the appendix. Below we lay out the conceptual approach. 

We assume that conditions describing a galactic body can be expressed as an index. The difference between the index values for any galactic body and the Earth defines an Earth-difference metric, $\theta$. A more precise definition for $\theta$, in terms of delineating all the physical/chemical factors that could feed into it, is not required for the analysis that follows. Figure 2 illustrates how $\theta$ is used in our thought experiments. The analysis is referenced to our own planet as $\theta_{Earth}$. We assume there is a range of conditions ($\theta_{min}$ to $\theta_{max}$) over which planetary life is possible. Galactic bodies are grouped into bins, denoted as $\theta_i$, each having a fixed width. Each bin holds an equal number of objects, an assumption that can be relaxed. Of those objects, some will be inhabited and some will not. Based on this premise, a probability index, characterizing the life potential of each set of conditions, is assigned to each bin. Effectively, we are assigning a habitability index, akin to that of Barnes et al. [2015], to each bin. The bin with the highest probability index is defined to be the set of conditions that maximizes life potential. The position of this bin, on the Earth-difference scale, is denoted by $\mu$. 

The procedure above allows a large number of distributions to be generated. As a starting point, we choose a normal distribution to model the frequency of inhabited planets based on their specified conditions. We then vary $\mu$ and create different probability distributions. Here, $\mu$ is varied between 0 and 0.5 at an increment of 0.1. Allowing the life potential peak to vary is resonant with Heller and Armstrong [2014] who have argued that Earth conditions may not be those that maximize life potential. To be clear, we are not arguing that this is or is not the case. Our stance is that at present we do not know. We consider it a possibility in the same sense that it is possible that Earth conditions do indeed maximize life potential. Effectively we are considering multiple working hypotheses and exploring the implications of each for different search strategies. 

Variable synthetic distributions, constructed as per above, represent different galactic potentials. The number of observations we have to date do not allow for discrimination between different potentials. For our experiments we will assume that future observations can be used toward this end and we will ask how different search strategies can achieve it. A search strategy is defined in terms of a search window centered about $\theta_{Earth}$ and extending to a value of $\theta_{Search}$ in both directions. The greater the extent, the more we are willing to search for inhabited, non Earth-like planets.  Increasing values of $\theta_{Search}$ are evaluated for there ability to recover the mean of the inhabited distribution, referred to as $\mu_{test}$, and to provide accurate estimates of galactic life.

Within this framework, we constructed different galactic life distribution potentials and considered different search strategies for each. For any distribution, we drew from it at random, subject to a specific search window, observing what conditions defined each object and whether they were inhabited or not. The number of observations needed to find a fixed number of inhabited objects was tracked. This number can be varied and we will also consider search strategy performance if the goal is to only find a single inhabited planet. If the goal is to provide some constraints on galactic life distribution then more than a single find would be required. The minimum number of observations needed to potentially constrain a distribution is not a fixed, agreed upon value. As a starting point, we settled on 30 inhabited objects - this number is only a `rule of thumb' for the minimum number of observations needed to potentially constrain a distribution [Hogg and Tanis, 1997]. Once this number of observations was obtained, the conditions that maximize life were estimated for varying search window widths (differences in the estimates could then be compared to actual distributions to gauge uncertainty as a function of search strategy). This process was performed a number of times, tracking the number of observations and life potential maximizing conditions that were arrived at each time. We then assessed the accuracy and cost of each search strategy. Accuracy was defined as the percent error between the conditions predicted by each search strategy and the true solution. Cost is related to the number of observations. The more observations needed, the greater the cost. Some function could be devised to approximate how this cost translates to total time and resources. Here we assume that the number of observations provides a useful starting point in considering relative costs (we appreciate that the scaling between number of observations and monetary cost will likely be a non-linear).

\begin{figure*}[!htb]
\includegraphics[width=\textwidth]{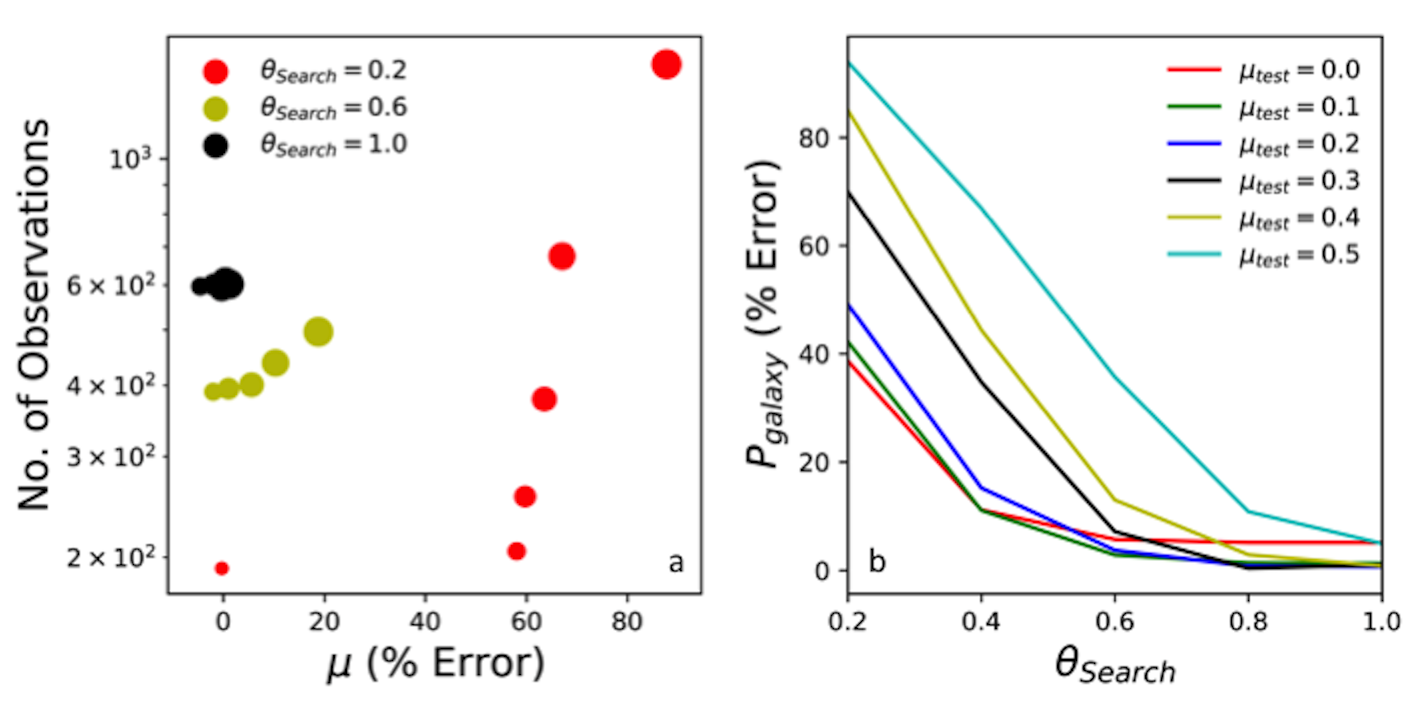}
\caption{a) A cost-benefit relationship based on search window width (color of dot) and how different the conditions can be from Earth and allow for life (larger dots are further from Earth-like). b) Accuracy is assessing how prevalent life is within the galaxy ($P_{galaxy}$) versus search window width for different potential galactic distributions ($\mu_{test}$).}
\end{figure*}

For the initial suites of experiments, the probability of life amongst galactic bodies at the distribution peak was set to 20\%. The probability then dropped toward zero as conditions moved away from the peak toward the most extreme cases that maintained life potential (Figure 2). This equated to assuming that roughly 5\% of galactic objects that reside within the window of planetary conditions that allow for life would be inhabited. That number can and will be varied extending to 25\% and less than 1\%. 

The first scenario considered was one under which the conditions that maximize life potential are centered on Earth conditions. In this case, a focused search strategy (i.e. a narrow search window) is ideal for finding a minimum number of inhabited planets in a cost efficient way (Figure 3a). It can under predict how prevalent life is in the galaxy by about a third (Figure 3b). Doubling the search window can bring the estimate of galactic life down to the lowest possible error at approximately twice the cost of a narrow search. 

If the conditions maximizing life potential are different from Earth then the accuracy and efficiency of different search windows changes (Figure 3). The changes for some search windows are large while for others they are relatively mild (i.e., different strategies have different levels of robustness). When the optimal conditions are similar to Earth ($\mu_{test}=0.1$), the error of the narrowest search strategy in estimating galactic life is not significantly increased (Figure 3b) and cost remains low (Figure 3a). The gains associated with moving to an intermediate search window are similar to the Earth centered case (Figure 3b). 

\begin{figure*}[!htb]
\includegraphics[width=\textwidth]{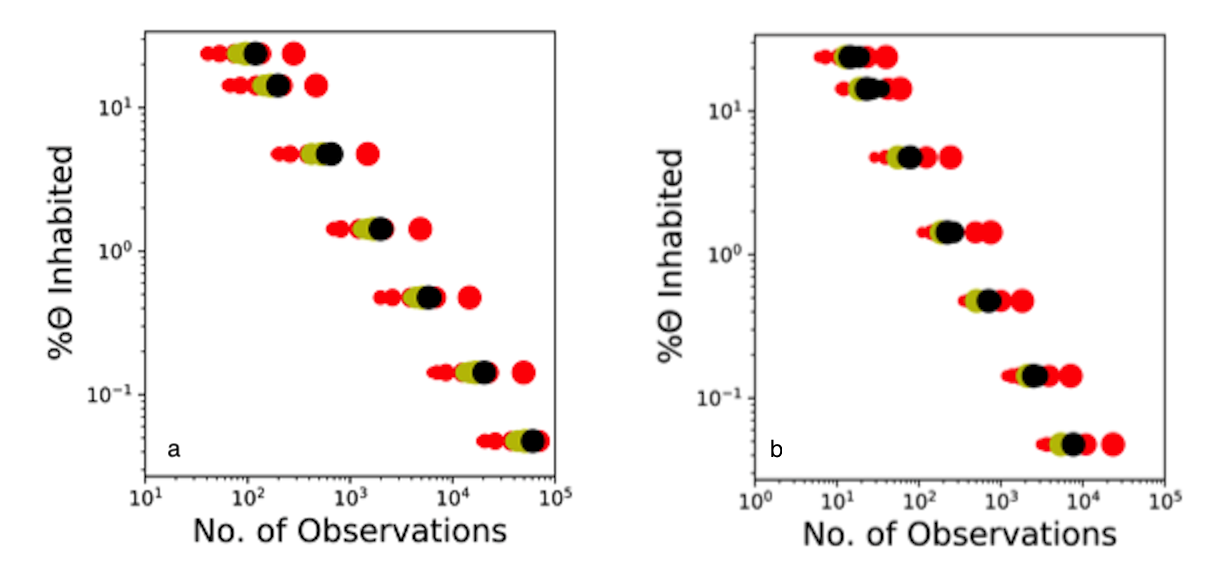}
\caption{Results from experiments, as per Figure 3, that vary the total percentage of inhabited galactic bodies.}
\end{figure*}

If conditions that maximize life potential deviate further from Earth ($\mu_{test}=0.3$) then changes in the cost-benefit relationships, for variable search strategies, can become significant. For the narrowest window, accuracy continued to decrease while cost, in terms of number of observations, increased relatively rapidly. The cost of a narrow search exceeded that of an intermediate search as $\mu_{test}$ exceeded 0.3. For those cases, a narrow search window leads to significant bias. As a result, a lot of observations (and time) are required and the prediction they lead to regarding galactic life potential is in significant error. This is a dangerous combination. For $\mu_{test}=0.3$, going from a narrow to an intermediate search can lower the error in assessing galactic life potential from over ~70\% to under 10\% with close to zero increase in cost. For $\mu_{test}>0.3$, going to an intermediate search can lower both error and cost. 

Thus far, we have focused on the contrast between narrow and intermediate search strategies. Extending the search to observe everything that we think is capable of harboring life increases accuracy over all potentialities. This comes with increased cost. Over most of the potentiality space we have explored the increased cost in going from an intermediate to a wide search does not come with a significant increase in accuracy. For the most extreme cases tested, there is a significant increase in accuracy. For those cases, the accuracy increase for a wide relative to an intermediate search is a factor of 4 with a factor of 1.25 increase in cost. In short, the principal gain in going from an intermediate to a wide search is that the risk, in terms of misrepresenting galactic life, for a ``worst-case'' situation is minimized. 

The absolute number of observations needed to find a fixed number of inhabited planets, for any search strategy, depends on the assumed percentage of inhabited galactic bodies. Figure 4 shows results from experimental suites that vary that value under the goal of finding 30 (Figure 4a) or a single inhabited planet (Figure 4b). For the experiments of Figure 4a, the accuracy in assessing galactic life follows the same trends as in Figure 3b. The relative difference between search strategies remains robust for different assumptions as to the total number of inhabited planets. The exception is that as the total number becomes very low the potential advantage of a wide search becomes weaker (Figure 4a, lower right corner). If the total number of inhabited planets is very low, then the number of planets that must be assessed to find more than a single inhabited planet becomes large under all search strategies. If that is the case, then we are pushing toward the limit at which statistical analysis is not justified. 

To this point we have assumed that finding a single inhabited planet is not the sole goal for explorations of galactic life. Nonetheless, it is useful to consider time commitment toward the first milestone within the overall search endeavor. The results of Figure 4b can be viewed in this way. They also have potential utility if the total number of inhabited planets in the galaxy is so low that efforts to determine a `life potential distribution' will be effectively doomed from the start. If finding a single inhabited planet is the sole goal and/or if it turns out that a very large number of planets need to be assessed before we find any signs of life on one of them, then Figure 4b is pertinent while issues of accuracy as per Figure 3b are less pertinent. There are some difference between a single planet and multi planet goal, particularly if the total number of inhabited galactic bodies is low. However, the main trend we are highlighting remains robust: a narrow search can be more efficient under some galactic life potentials but it comes with greater risk when evaluated over a broad range of potentials that extend beyond the assumption that Earth conditions maximize life potential. 

\subsection*{Changing Minds}

Different ideas about life in our galaxy (e.g., rare versus plentiful) can be viewed as competing hypotheses or, in a probabilistic framework, different \emph{a priori} assumptions that have not been fully tested (in this case because observations are currently limited). Each end-member idea is a viable hypothesis. Each is testable. That is, each is a scientific prior. Any scientific prior will adjust to new observations and, given a large number of observations, all such priors should converge toward the hypothesis that is most consistent with observations. In principal, it does not matter what prior assumptions a search strategy is based on - the observations will decide in the end. In practice, we need to consider that the observations we will make over the next wave of missions will be discrete in number. Using the Earth-likeness metric of the previous sub-section we can ask which hypothesis is more responsive to discrete new observations. 

The discussion above relates to search strategies. One can imagine that a scientist who holds to a rare-Earth, or Earth-centered, view would opt for a narrow search window on the grounds that this would be the most efficient strategy. On the other hand, a scientist who holds to the view that a wide range of conditions can allow for planetary life might argue against a narrow search on the grounds that it will introduce a bias. We can pose the question of how many discoveries of inhabited planets would be required to change either scientist's mind if their hypothesis happens to be incorrect. 

\begin{figure}[!htb]
\includegraphics[width=0.5\textwidth]{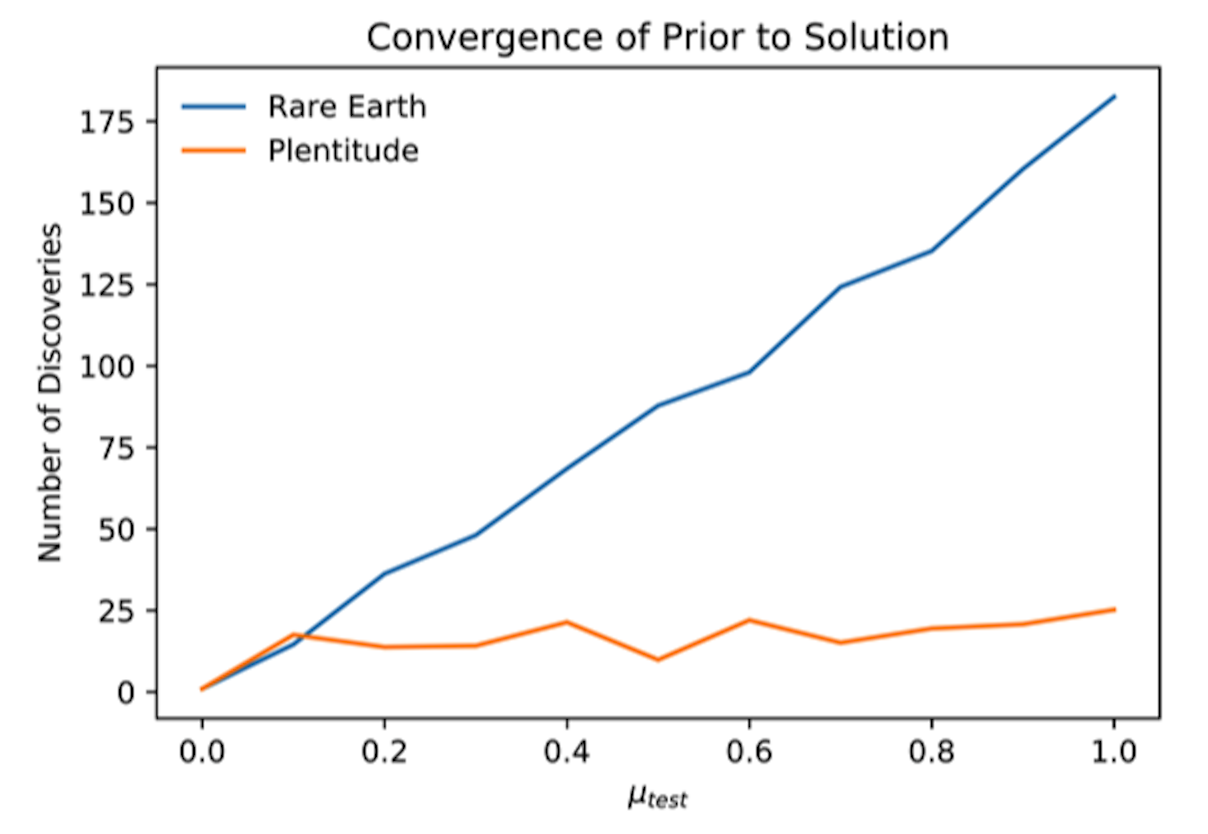}
\caption{The number of discoveries, of inhabited planets, it takes to converge to the attributes that maximize inhabitance for different priors and for different life probability potentials.}
\end{figure}

The question above may seem highly subjective. However, asking the question of how \emph{\emph{a priori}} ideas or beliefs adjust to new observations lends itself to a Bayesian analysis (Appendix), which can provide quantitative insight. For an Earth-centered prior, initial confidence begins tightly peaked around $\theta_{Earth}$ . A plentitude prior assigns equal weight across some $\theta$-range. As in our previous analysis, we can draw observations randomly and track the conditions of the body and whether or not it is inhabited. If the body is indeed inhabited, both end-member priors are updated to reflect this. This is repeated and the number of observations tracked until each scientist's confidence is in alignment with the true set of conditions that maximize life potential within an equal tolerance. This is quantified using a maximum a posterior (MAP) metric [Robert, 2002]. That metric is defined as the parameter with the maximum posterior probability (further details can be found in the appendix). 

Figure 5 shows how long it takes each end-member hypothesis to converge for different potential galactic life distributions. If galactic conditions are Earth centered, it would not take long to alter either \emph{\emph{a priori}} view. If the conditions are different than Earth, it takes more observations for both priors to converge to the true solution. This increase occurs because our first data point is always Earth. That initial data constraint is a restatement of our current understanding of life in the galaxy: we have only observed one inhabited body, Earth. If conditions maximizing life potential are far different from Earth, then our planet can introduce an initial bias. A plentitude view adjusts relatively quickly, regardless of our planets galactic status. An Earth-centered prior has a slower adjustment in the face of new observations, i.e., it comes with a greater resistance. If galactic life potential is not Earth centered, then going in with the assumption that it is, as opposed to the assumption that a wider range is possible, will lead to $\sim$6 times more observations being required to remove the shadow of the \emph{\emph{a priori}} assumption.  


\section*{Discussion}

In the previous section we used an Earth-likeness (equivalently a nonEarthness) metric to evaluate the cost, accuracy, and risk associated with different search strategies and the more subjective cost associated with competing \emph{a priori} hypothesis regarding planetary life potential. We employed a generic metric to provide a quantitative measure of the degree of nonEarthness that maintains life potential. In the statistical thought experiments, the results were dependent on the value of this metric to variable degrees - in some scenarios that cannot be observationally ruled out the degree of dependence was large. An upshot is the suggestion that there is value in developing, as a community, a more tightly defined metric. This could provide some nonEarth1.0 balance to Earth2.0 thinking and discussions. The balance could have scientific and humanistic implications. In the following sub-sections, we wade into these discussion topics, not as a final say, but as a starting point.

\subsection*{Different is More}

What is the value of considering the potential of inhabited, nonEarth-like planets in the search for life? What is the value of finding such a planet? The first question was at the core of the previous section. A key implication was related to risk minimization, which we expand on below. We then move to the second question. We focus in on Gaia theory and on how exoplanet exploration could provide a useful Gaia test bed that extends beyond Earth. 

Search strategies should be able to achieve goals in cost efficient ways that minimize risk in the face of worst-case scenarios. What is the worst-case scenario for investing resources to assess galactic life potential? Is it finding no signs of life beyond Earth? That is a worst case if one believes that it is a negative conclusion. If the goal is to reach a conclusion we can be as confident of as we can be, then no conclusion is negative if it is the one that follows from the preponderance of un-biased evidence. We would argue that the worst case would be spending a large amount of time and resources to reach the wrong conclusion. 
Figure 3 indicates that an overly narrow search can open us to that risk. 

The above can be viewed as an overly ``formalized'' assessment. Exploration depends on public interest, which doesn't fit as easily into formalized analysis. From that angle, a worst-case scenario could be viewed as one that tries public patience (although we would argue that an equally valid worst-case scenario would be misinforming the public). This could motivate the idea that we need to find signs of an inhabited planet beyond Earth as soon as we can (the opening statements quoted in the introduction reflect that, rightly or wrongly, this has already been effectively promised to the public). This, in turn, could be used to argue for an Earth-centered search based on the idea that it would minimize time commitment. The difficulty is that we would then be making a time investment decision based on assuming we know the answer to a critical question before we put the investment strategy into practice. This again opens us to risk as a narrow search could come with a higher time cost if we are wrong in terms of our \emph{a priori} assumption (Figure 4b).

In regards to assumptions, all of our results required assumptions regarding galactic life distributions. This, we would argue, does not weaken our conclusions but is an overarching point of the analysis to begin with. The tighter a search focuses around an Earth2.0 mode, the more a soft form of rare-Earth is being assumed to be valid (i.e. life potential peaks at Earth). One could argue that models of planetary habitability and observations from our solar system, to date, lend support to this assumption. Fair enough but it remains an assumption. It may prove correct but if it is used to justify a search strategy then the strategy is designed, effectively, to test only one hypothesis. One could take an alternate view. Rather than arguing for one hypothesis over another (e.g., life potential peaking near or away from Earth [e.g., Heller and Armstrong, 2014]), we can assume that a range of hypothesis remain viable and seek to minimize risk in the face of the various potentialities. This is the view we have argued for. Considering the potential of inhabited non-Earth like planets can minimize risk at an increased cost that, in our opinion, is not excessive (Figures 3 and 4). It also allows for the potential of discriminating between different hypotheses regarding galactic life (Figure 5). 

We now turn to the question of ``what is the value of finding an inhabited planet that is non Earth like in other regards?'' The classic concept of a ``Habitable Zone'' assumes that delineating planetary conditions that allow for the potential of life can proceed without considering life's role [Kasting et al., 1993]. In effect, it is assumed that habitability can be determined without explicitly considering inhabitance. This assumption has been challenged and the degree to which removing it from exoplanet discussions could influence our thinking about life in our galaxy is significant [Goldblatt, 2016]. The difficulty has been and remains that observations from this planet can not unravel the degree to which life influences cycles that regulate environmental conditions (the Earth has life and remains geologically active - which of these is more critical to the Earth being habitable is difficult to unravel as life has entwined itself in a range of geophysical/geochemical cycles [Goldblatt, 2016]).  

From an exoplanet perspective, we can push Gaia theory to a limit. If a strong form of Gaia can operate then planetary regulation could occur without abiotic cycles (Figure 1c). That is, life could do the heavy lifting for maintaining habitable conditions. The implication is that although planetary internal energy, that drives volcanism and tectonics, plays a role for the Earth's inhabitance it is not critical for planetary bodies in general. The internal energy of a planet will depend on its age and composition. The potential of determining composition and age for exoplanets is actively being discussed within the community. Both could be within reach for next generation observations. If a planet is found that has low potential of being geologically active and shows biosignatures this would provide a step toward confirming Gaia (internal energy may still have been crucial for the origin of life [Baross and Hoffman, 1985] and for maintaining habitable conditions early in the planets lifetime but the extension of habitable conditions past the geologic lifetime of the planet would be due to life itself). At present, search strategies are focused on planets that are likely to be geologically active with the thought that this is critical for life [Ward and Brownlee, 2000; Kasting, 2010]. That remains an assumption. It is an assumption that has the potential to be refuted if a single strong Gaia1.0 is found (a greater number of Earth2.0's would be required to confirm the assumption at a statistical level).  

Finding strong Gaia1.0 would dramatically change our views about planetary habitability (finding Earth2.0 would be a confirmation of a prevalent idea). The degree of rethinking can be hinted at by posing a question: Is habitability a characteristic of a planet, like temperature, or is it something that flows through it, like heat? Stated another way, is it a state or a process variable? The classic "habitable zone" concept assumes that it can be treated as a state variable. From that, follows the assumption that its limits can be determined so as to effectively make a phase diagram that delineates regions that allow for life. On a strong Gaia the origin and evolution of life are dominant for habitability. With that comes contingency and the potential of multiple temporal paths leading to variable end states of inhabitance [Walker et al., 2018]. This is a process variable view [Bridgman, 1943] under which multiple equilibrium states are possible and path-dependence cannot be ignored [Dyke and Weaver, 2013; Weaver, 2015; Lenardic et al., 2016]. This brings in layers of potentiality associated with evolution and historical contingency (to be clear, a contingent process is not the same as a random process [e.g., Bohm, 1957]). The approach to planetary life research would, as a result, need to move toward one that is more statistical/probabilistic than it is at present [Walker et al., 2018].  

The issue of evolution leads to a final point. Rare-Earth ideas acknowledge that planets different from the Earth could have simple (microbial) life but argue that higher life (plants and animals) requires conditions like that of Earth [Ward and Brownlee, 2000]. Life can respond to environmental changes. Many of the changes, that are considered critical to the development of higher life on Earth, are ascribed to internal energy sources driving changes in surface conditions such that if a planet lacked the geological changes that occurred on Earth it could have microbial life but it would not have developed higher life [Ward and Brownlee, 2000; Stern 2016]. The idea that evolution requires environmental changes is not agreed upon for the evolution of life on Earth [McKee, 2000] and, as such, extending it to planetary bodies in general is an \emph{\emph{a priori}} assumption. Exoplanet search strategies have incorporated the potential that atmospheric bio-signatures might be of the kind that prevailed on early Earth, before the rise of complex life, or of the kind associated with higher life [Schwieterman et al., 2018]. Finding signs of higher life on a geologically inactive planet could provide a new layer of evidence that evolution can proceed in an autocatylistic mode with no need for externally driven environmental changes [Kauffman, 1993; McKee, 2000; Cazzolla Gatti, 2011]. 

Finding signs of life on another planet, be it like our own or significantly different, would be a major discovery. The implications of that discovery could be further reaching for planets that are different. In that sense, different is more - it could bring more information content about life potential in our galaxy.

\subsection*{Narratives}

The search for life beyond Earth is no small undertaking. It can benefit from efforts to engage the public. The engagement is often framed as a narrative. Narratives can go well beyond public relations. A narrative can frame a problem in a way that favors one decision/conclusion over alternatives that are just as rational as the frame-favored decision [Tversky and Kahneman, 1981]. For the issue at hand there is a feedback as public opinion can influence exploration strategies. Whether it is intended or not, building a narrative around a problem will influence the way humans think about the problem and, in keeping with the cultural/humanistic connections of this sub-section, `what we think changes how we act' [c.f., Gang of Four, Solid Gold, EMI/Warner Bros., 1981]. In short, this encapsulates the value of re-framing problems and considering multiple frames when addressing problems involving uncertainty [Tversky and Kahneman, 1986]. 

An Earth2.0 narrative that has been used to frame exoplanet exploration reinforces the idea that Earth conditions are the ideal ones for habitability (a variant of a rare-Earth narrative [Ward and Brownlee, 2000]). We would argue that we do not have the observations needed to discriminate between different assumptions regarding galactic life potential at this stage of our exploration and it is not in the best interest of the search to send messages, explicit or implicit, that we do. An Earth2.0 narrative, as it is being presented to the public, walks the line of promising specific returns and the dangers of that for science, in the public realm, should be kept in mind [e.g. Riordan et al., 2015].

Beyond sending messages that do not accurately reflect the state of our knowledge, an Earth2.0 narrative comes with philosophical and cultural baggage that may not best serve its intended purpose. An Earth2.0 narrative is nostalgia heavy. There have been many words put forward in the service of an Earth2.0 narrative but images, arguably, can give a better sense of the messages this narrative carries. Artwork depicting travel posters to exoplanets with white picket fences [https://exoplanets.nasa.gov/alien-worlds/exoplanet-travel-bureau/] invoke a sense of the familiar - a sense of home (note: only a sense of home for those who grew up culturally and economically in places that had white picket fences to begin with). The desire for a place just like home is fed by nostalgia and a desire for an ideal [Messeri, 2016] - an ideal that may never have existed. The potentially false narrative of an ideal past has been taken up in artistic works (e.g., some movie examples: Plaesantville, Midnight in Paris, Trainspotting 2) and in historical studies [e.g., Webb, 2017]. We mention these works to point out that, to a portion of humanity, a nostalgia filled narrative may not have the effects hoped for.

We are not implying that the cultural value of nostalgia in general, or specifically for developing public narratives, can be broken down to 1's and 0's. It is not as simple as it's always good or always bad. It depends on context [e.g. Bonnett, 2010]. Within the context of space exploration, a narrative built on finding a second Earth is intertwined with the idea of finding a second home [Messeri, 2016]. Home is something we know, something comfortable, and ideas of home can lead to thoughts of "better times". This can send the message that our goals are to recapture something. It invokes a sense of looking toward the past, which runs counter to the idea of exploration. When pushed to limits, such nostalgic messages are associated with populist movements. From our perspective, this also runs counter to the idea of space exploration, which is a global endeavor (i.e., an internationalist as opposed to a national endeavor).

An alternate narrative, as compared to searching for an Earth2.0, is one of galactic diversity in terms of livable and living planets. In this alternative narrative, the future becomes more prominent with all the lack of certainty, immediate comfort and familiarity that a future holds. Planets with life beyond Earth may be something foreign to us. They may be uncomfortable for us to live on at first. To know them we need to find them, as opposed to starting our exploration on the premise we already know them (we are not the first to remark on the dangers of assuming we know the answer before hand when it comes to planetary exploration [Moore et al., 2017; Tasker et al., 2017]). The alternate framework we are proposing will also not resonate across all of humanity as a public engagement narrative but which of the two, diversity of living planets versus finding a second Earth, better represents the sense of exploration that got us, as human beings, to begin exploring space in the first place?


\section*{Appendix: Methods}

\subsection*{Statistical Cost-Benefit-Risk Analysis}

We used a statistical analysis to address our motivating question: How does search window width affect the cost-benefit relationship based on the conditions that maximize life potential?  In the analysis, we tested three search window widths - narrow, intermediate and wide, which are defined by the $\theta_{Search}$ values 0.2, 0.6 and 1.0, respectively. For each search window we tested six different life distributions, each having increasingly nonEarth-like life potential maximizing means ($\mu_{test}$), ranging from 0.0 to 0.5. For each combination of $\theta_{Search}$ and $\mu_{test}$, we conducted 100 trials that randomly sampled a synthetic data set, characterized by the conditional probability P(I|$\Theta$), the probability that an object is inhabited given a specific set of conditions. We assumed that each subset of conditions, broken up into discrete intervals ($\theta_i$), 0.1 units in width, were equally likely to represent any observable object. Within each $\theta_i$ interval, some of the objects were inhabited and some were not. The percentage of inhabited bodies in each $\theta_i$ interval was approximated as the probability for that bin given a normal distribution defined by $\mu_{test}$ and a standard deviation ($\sigma$) of 0.2, which we held constant. 

For each combination of $\theta_{Search}$ and $\mu_{test}$, a set of trials were performed to estimate the sample statistics. In each trial, we used a uniform random number generator to determine the bin from which an observation would be drawn (subject to limits imposed by the search window). Then, we did a second random draw, from that bin, and tracked whether the observation was of an inhabited or uninhabited object (the greater the habitability probability from that bin, the greater the chances that an inhabited object would be observed). During the duration of the trial, we tracked the total number of observations, the conditions that defined each observed object, how many objects were accumulated in each interval $\theta_i$ and whether or not each object was inhabited. Once 30 inhabited objects were observed, we calculated the mean $\bar{x}$ and standard deviation (\emph{s}) of the sample. Following the 100 different, randomized trials, we calculated the means $\mu_{\overline{x}}$, $\mu_s$, $\mu_{n_i}$ and $\mu_{n_{\theta_i}}$ of the sampled statistics. In this paper, we refer to $\mu_{\overline{x}}$ as the observed mean, $\mu_s$ as observed standard deviation, $\mu_{n_i}$ as the number of observed inhabited objects in the interval $\theta_i$ and $\mu_{n_{\theta_i}}$ as the total number of observations in the interval $\theta_i$. 

Using the results of the statistical analysis, we calculated the percent error between the true ($P_{galaxy}$) and predicted number of inhabited galactic objects ($P_{galaxy}^*$) for each combination of $θ_{Search}$ and $μ_{test}$. The statistical analysis results tell us the average ratio between the number of inhabited objects and total observed objects is defined as

\begin{equation}\label{fraction1}
\frac{\mu_{n_i}}{\mu_{n_{\theta_i}}}
\end{equation}

for each $\theta_i$. Equation \ref{fraction1} can be scaled to the entire population by the relation

\begin{equation}\label{fraction2}
\frac{\mu_{n_i}}{\mu_{n_{\theta_i}}}=\frac{N_i}{N_{\theta_i}}
\end{equation}

where $N_i$ is the total number of inhabited objects in the interval and $N_{\theta_i}$ is the total number of objects in the interval. Using our mean of the distribution of the means $(\mu_{\bar{x}}$ and $\mu_s)$, we can approximate the probability distribution of life conditions as 

\begin{equation}\label{Prob1}
P(I|\Theta)=N(\mu_{\bar{x}},\mu_s)
\end{equation}

Furthermore, for each interval, we know that 

\begin{equation}\label{Prob2}
P(I|\theta_i)=\frac{N_i}{P_galaxy},
\end{equation}

where $P_{galaxy}$ is the number of objects in the galaxy that are inhabited. Combining equations \ref{fraction1}, \ref{fraction2} and \ref{Prob2}, we arrive at our estimate of the galactic population

\begin{equation}\label{Pgalstar}
P^*_{galaxy}=\frac{N_{\theta_i}n_i}{\mu_{n_{\theta_i}}P(I|\theta_i)}.
\end{equation}

Following the same procedure, except using $\mu_{test}$ and $\sigma$ rather than $\mu_{\bar{x}}$ and $\mu_s$ in equation \ref{Prob1}, the true galactic population is

\begin{equation}\label{Pgal}
P_{galaxy}=\frac{N_{\theta_i}n_i}{\mu_{n_{\theta_i}}P(I|\theta_i)}
\end{equation}

where $n_i$ is the true number of bodies present in the interval \emph{i} if a representative sample were taken of size $\mu_{n_{\theta_i}}$. That is 

\begin{equation}\label{ni}
n_i=\mu_{n_{\theta_i}}P(I|\theta_i).
\end{equation}

Substituting equation \ref{ni} into \ref{Pgal} we get 

\begin{equation}\label{Pgal2}
P_{galaxy}=N_{\theta_i}.
\end{equation}

This relationship is a result two main assumptions: (1) that every object is equally likely to be characterized by any set of conditions $\theta_i$, and (2) the proportion of objects that are inhabited within a given interval \emph{i} is represented as the probability of that interval for the normal distribution characterized by $\mu_{test}$ and $\sigma$. To obtain the percent error ($\epsilon$), we use the relation 

\begin{equation}\label{eps1}
\epsilon = \left|{\frac{P_{galaxy}-P^*_{galaxy}}{P_{galaxy}}}\right|,
\end{equation}

which, after combining with equations 5 and 8, simplifies to

\begin{equation}\label{eps2}
\epsilon = \left|1-\frac{\mu_{n_i}}{\mu_{n_{\theta_i}}P(I|\theta_I)}\right|.
\end{equation}

\subsection*{Changing Minds}

Bayesian inference provides a framework to evaluate how new data can alter \emph{\emph{a priori}} assumptions that define different end-member search strategies. Assumptions regarding our current state of knowledge are used to assign probabilities to a particular set of beliefs known as the prior, $P(\theta)$, where $\theta$ is a discrete vector representing parameters. As new data (\emph{D}) are discovered, the chances of them occurring, given that a particular parameter is correct, is then computed to produce the likelihood function, $P(D|\theta)$. The likelihood is telling us the probability of getting the data given the set of conditions. The prior and likelihood function are then combined to produce our updated knowledge, known as the posterior $P(\theta|D)$, which is interpreted as the probability that parameter $\theta$ is correct, given the observed data set \emph{D}. These three components are related by 

\begin{equation}\label{bayes1}
P(\theta|D) \propto P(D|\theta)P(\theta)
\end{equation}

In our analysis, $\theta$ represents how Earth-like a body is and \emph{D} represents the number of such galactic bodies that are inhabited. Earth-likeness is a sliding parameter scale normalized to Earth at a value of zero and extending to negative and positive infinity. We consider a finite range between some assumed minimum and maximum value, $\theta_{min}$ and $\theta_{max}$ ($\theta_{min}$ and $\theta_{max}$ are assigned normalized values of -1.0 and 1.0, respectively, for the most extreme version of a plentitude hypothesis). We consider differing values of Earth-likeness at intervals of 0.1, totaling 21 different types of Earth-likeness bins. 

We use different priors to represent competing hypotheses for the range of earth-likeness an inhabited planet can have. One prior assigns a high probability around zero in $\theta$-space and much lower values everywhere else. It is represented as a normal distribution with a mean value of zero and variance of 0.1. The second prior assumes that all earth-likenesses, within our finite range of consideration, are equally probable to be inhabited. 

Discoveries of future hypothetically inhabited planets are used to produce the likelihood function. It is assumed that the Earth-likeness of each new discovery can be categorized. Using this data, the standard Gaussian functional form is used to derive the likelihood function, 

\begin{equation}\label{likelihood}
P(D|\theta)=\prod_{i=1}^N\frac{1}{\sqrt{2\pi}\sigma}exp\left[-\frac{(\theta-D_i)^2}{2\sigma^2_i}\right]P(\theta)
\end{equation}

where we assume each new data point is independent of any other, potentially a limiting assumption as the sampling may be highly selective. This allows for the multiplication of the prior and likelihood function to produce the posterior. The posterior is then normalized such that the sum of probabilities from all bins equal one.

A plentitude hypothesis uses an uninformative prior, which allows the data to directly influence the posterior distribution. The posterior is computed by calculating the likelihood function and normalizing the sum of probabilities to one. On the other hand, an Earth-centered prior is normally distributed and is multiplied with the likelihood function producing a posterior that is also normally distributed. The mean and variance of this posterior is computed as

\begin{align}
a&=\frac{1}{\sigma^2_{prior}}, \\
b&=\frac{n}{\sigma^2}, \\
\mu_{post}&=\frac{a\mu_{prior}+b\bar{x}}{a+b}, \\
\sigma^2_{post}&=\frac{1}{a+b}
\end{align}

where $\sigma$, $\bar{x}$ and \emph{n} are the known uncertainty of the likelihood, the mean value of the data and number of data points, respectively. Once the probabilities have been computed, they must be normalized to sum together to one.

To gauge the convergence of a given prior we use the maximum a posteriori (MAP), defined as the parameter, which has a maximum posterior probability [Robert, 2002]. The MAP is used to calculate the convergence of end-member priors to a final solution. The number of discoveries it takes for the MAP to equal the assumed mean of $\theta$, given random draws from an assumed data set, defines the ``convergence metric'' plotted in Figure 5. 

To test the convergence of each prior, we assume a normally distributed subset of 1,000 inhabited planets (other values can be tested - our initial interest is a relative comparison between two end-member priors and, as such, absolute numbers for convergence are not particularly meaningful as they are overly dependent on the assumed number of inhabited bodies). The mean of different potential galactic distributions ($\mu_{test}$) is varied between 0 and 1 in $\theta$-space with a standard deviation of 0.3. Planets are drawn at random, following Earth as the initial data point, and the number of draws it takes to converge to a solution is tracked. The process is repeated for each $\mu_{test}$ and each prior 150 times and the average number of draws, from the subset of 150, is reported as the convergence metric for each prior under each variable galactic life potential (Figure 5).


\section*{References}

\begin{small}

Abe, Y., A. Abe-Ouchi, N.H. Sleep, and K.J. Zahnle (2011), Habitable Zone Limits for Dry Planets, Astrobiology, 11, doi: 10.1089/ast.2010.0545  \\

\noindent Abbot, D. S. and Switzer, E. R. (2011), The Steppenwolf: A Proposal for a Habitable Planet in Interstellar Space, Astrophys. J. Lett., 735(2):L27.  \\

\noindent Barlow, C. - Editor (1991), From Gaia to Selfish Genes: Selected Writings in the Life Sciences, MIT Press, ISBN: 9780262521789.  \\

\noindent Baross, J.A., and Hoffman, S.E. (1985), Submarine hydrothermal vents and associated gradient environments as sites for the origin and evolution of life, Orig. Life Evol. Biosph., 15, 327-345. \\

\noindent Barnes, R., B. Jackson, R. Greenberg, and S. Raymond (2009) Tidal limits to planetary habitability, Astrophys. J., 700, L30, doi:10.1088/0004-637X/700/1/L30.    \\

\noindent Barnes, R., V.S. Meadows, and N. Evans (2015) Comparative habitability of transiting exoplanets, Astrophys. J., 814:91, doi:10.1088/0004-637X/814/2/91.    \\

\noindent Bean, J.L., D.S., Abbot, and E.M.-R. Kempton (2017), A statistical comparative planetology approach to the hunt for habitable exoplanets and life beyond the solar system, Astrophys. J. Lett., 841, L24.  \\

\noindent Berner, R. A., A. C. Lasaga, and R. M. Garrels (1983), The carbonate-silicate geochemical cycle and its effect on atmospheric carbon dioxide over the past 100 million years, Am. J. Sci., 283, 641-683.   \\

\noindent Bohm, D. (1957), Causality and Chance in Modern Physics, London, UK: Routledge and Kegan Paul.  \\ 

\noindent Bonnett, A. (2010), Left in the Past: Radicalism and the Politics of Nostalgia, ISBN: 978-0826430076 Bloomsbury Academic, London.    \\

\noindent Bridgman, P.W. (1943), The Nature of Thermodynamics, Harvard University Press.  \\

\noindent Cazzolla Gatti, R. (2011), Evolution is a cooperative process: the Biodiversity-related niches differentiation theory (BNDT) can explain why, Theor. Biol. Forum 104(1), 35-44.    \\

\noindent Couprie, D.L. (2011), Heaven and Earth in Ancient Greek Cosmology From Thales to Heraclides Ponticus, Springer, New York.    \\

\noindent Dyke J.G. and I.S. Weaver (2013), The emergence of environmental homeostasis in complex ecosystems, PLoS Comput Biol 9(5): e1003050, doi:10.1371/journal.pcbi.1003050.    \\

\noindent Foley, B.J., and A.J. Smye (2018), Carbon cycling and habitability of Earth-sized stagnant lid planets, Astrobiology, 18:7, 873-896, https://doi.org/10.1089/ast.2017.1695. \\

\noindent Goldblatt, C. (2016), The inhabitance paradox: how habitability and inhabitancy are inseparable, NASA Conference Publication, https://arxiv.org/abs/1603.00950.    \\

\noindent Heller, R., and J. Armstrong (2014), Superhabitable worlds, Astrobiology, 14:1, 50-66, https://arxiv.org/abs/1401.2392.  \\

\noindent Hogg, R.V., and Tanis, E.A. (1997), Probability and Statistical Inference, Prentice Hall, Upper Saddle River, NJ.   \\

\noindent Kaltenegger L,  and Sasselov D. (2011), Water-Planets in the Habitable Zone: Atmospheric Chemistry, Observable Features, and the case of Kepler-62e and -62f, Astrophys. J., 736, L25, doi:10.1088/2041-8205/736/2/L25.    \\

\noindent Kasting, J. (2010), How to Find a Habitable Planet, by James Kasting, Princeton University Press, Princeton, NJ.    \\

\noindent Kasting, J.F., Whitmire, D.P., and Reynolds, R.T. (1993), Habitable zones around main sequence stars, Icarus 101, 108-128.    \\

\noindent Kauffman, S.A. (1993), The Origins of Order, Oxford University Press.    \\

\noindent Kirchner, J.W. (1989), The Gaia hypothesis: Can it be tested?,  Reviews of Geophysics, 27, doi:10.1029/RG027i002p00223.    \\

\noindent Kirchner, J.W. (2003), The Gaia Hypothesis: Conjectures and Refutations, Climatic Change, 58, doi:10.1023/A:1023494111532.    \\

\noindent Kite, E.S., and E.B. Ford (2018), Habitability of exoplanet waterworlds, Astrophys. J., Submitted (https://arxiv.org/abs/1801.00748).    \\

\noindent Kite, E.S., E. Gaidos, and T.C. Onstott (2018), Valuing life-detection missions, Astrobiology, 18(7), DOI: 10.1089/ast.2017.1813.  \\

\noindent Kump, L. E., S. L. Brantley, and M. A. Arthur (2000), Chemical weathering, atmospheric CO2, and climate, Annu. Rev. Earth Planet. Sci., 28, 611-667.    \\

\noindent Lenardic, A., J.W. Crowley, A.M. Jellinek, and M.B. Weller (2016), The solar system of forking paths: bifurcations in planetary evolution and the search for life bearing planets in our Galaxy, Astrobiology 16 (7), http://dx.doi.org/10.1089/ast.2015.1378.   \\

\noindent Lenardic, A, A.M. Jellinek, B. Foley, C. O'Neill, and W.B. Moore (2016b) Tectonic-Climate Coupling: Variations in the Mean, Variations about the Mean, Variations in Mode.  J. Geophys. Res. Planets, doi:10.1002/2016JE005089.    \\

\noindent Lingam, M., and A. Loeb (2018a) Optimal target stars in the search for life, Astrophysical Journal Letters, 857:2.  \\

\noindent Lingam, M., and A. Loeb (2018b) Subsurface exolife, International Journal of Astrobiology, https://doi.org/10.1017/S1473550418000083.  \\

\noindent Lovelock, J. (1979) Gaia: A New Look at Life on Earth, Oxford University Press, Oxford, UK.  \\  

\noindent Lovelock, J. (1995) The Ages of Gaia: A Biography of Our Living Earth, Oxford University Press, Oxford, UK.    \\

\noindent Lovelock, J. E., and L. Margulis (1974), Atmospheric homeostasis by and for the biosphere: The Gaia Hypothesis, Tellus, 26, 1-10.    \\

\noindent McKee, J.K. (2000), The Riddled Chain: Chance, Coincidence, and Chaos in Human Evolution, Rutgers University Press.     \\

\noindent Messeri, L. (2016), Placing Outer Space: An Earthly Ethnography of Other Worlds, Duke University Press.  \\

\noindent Moore, W.B., A. Lenardic, A.M. Jellinek, C. Johnson, C. Goldblatt, and R. Lorenz (2017), How Habitable Zones and Super-Earths Lead Us Astray, Nature Astronomy, 1, doi: 10.1038/s41550-017-0043.    \\

\noindent Riordan, M., L. Hoddeson, and A.W. Kolb (2015), Tunnel Visions: The Rise and Fall of the Superconducting Super Collider, The University of Chicago Press, Chicago.    \\

\noindent Robert, C.P. (2002), The Bayesian Choice: from Decision-Theoretic Foundations to Computational Implementation, Springer, 2002.    \\

\noindent Schultze-Makuch, D. and L.N. Irwin (2001), Alternative Energy Sources Could Support Life on Europa, Eos, v82, No. 13, 150-151.    \\

\noindent Schultze-Makuch, D., A. Mendez, A.G. Fairen, P. von Paris, C. Turse, G. Boyer, A.F. Davila, M.R. de Sousa Antonio, D. Catling, and L.N. Irwin (2011), A two-tiered approach to assessing the habitability of exoplanets, Astrobiology, v11, No. 10, DOI: 10.1089/ast.2010.0592.    \\

\noindent Schneider, S., J.R. Miller, E. Crist and P.J. Boston - Editors (2008), Scientists Debate Gaia: The Next Century, MIT Press, Cambridge.    \\

\noindent Schwieterman, E.W., and 17 others (2018), Exoplanet biosignatures: a review of remotely detectable signs of life, Astrobiology, in press.   \\

\noindent Stern, R. (2016), Is plate tectonics needed to evolve technological species on exoplanets?, Geoscience Frontiers, 7, 573-580.     \\

\noindent Tasker, E., J. Tan, K. Heng, S. Kane, D. Spiegel and the ELSI Origins Network Planetary Diversity Workshop (2017), The language of exoplanet ranking metrics needs to change, Nature Astronomy 1, 0042, DOI: 10.1038/s41550-017-0042.  \\

\noindent Tversky, A., and D. Kahneman (1981), The framing of decisions and the psychology of choice, Science 211, 453-458.     \\

\noindent Tversky, A., and D. Kahneman (1986), Rational choice and the framing of decisions, The Journal of Business 59, S251-S278.  \\

\noindent Walker, J.C.G., P.B. Hays and J F Kasting (1981) A negative feedback mechanism for the long-term stabilization of Earth's surface temperature. J. Geophys. Res. 86, 9776-9782.    \\

\noindent Walker, S.I, and 13 others (2018), Exoplanet biosignatures: future directions, Astrobiology, in press.   \\

\noindent Ward, P.D. and Brownlee, D. (2000), Rare Earth, Copernicus Books.    \\ 

\noindent Watson, A. J., and J. E. Lovelock (1983), Biological homeostasis of the global environment: The parable of Daisyworld, Tellus B, 35, 284-289.     \\

\noindent Weaver I.S. (2015), Macroscopic Principles for the Self-Organisation of Complex Ecoystems, PhD Thesis, University of Southampton.    \\

\noindent Webb, S. (2017), Post-War Childhood: Growing up in the not-so-friendly Baby Boomer Years, Pen and Sword History, London.     \\

\noindent Wenz, J. (2017), Overlooked ocean worlds fill the outer solar system, Scientific American.  \\

\end{small}


\end{document}